\documentstyle[11pt,aaspp4]{article}
\input psfig.sty
\newcommand{\beq}{\begin{equation}} 
\newcommand{\eeq}{\end{equation}} 
\newcommand{\etal}{et~al.~} 
\newcommand{\fa}{$f(\alpha\!)~$}
\newcommand{\al}{$\alpha\:$}
\lefthead{Chappell and Scalo}
\righthead{Multifractal scaling in the ISM}

\input psfig.sty
\begin{document}

\title{Multifractal Scaling, Geometrical Diversity,\\
and Hierarchical Structure\\
in the Cool Interstellar Medium}

\author{David Chappell and John Scalo}
\affil{Astronomy Dept., University of Texas, Austin, TX 78712}

\begin{abstract}
	Multifractal scaling (MFS) refers to structures that can be 
described as a collection of interwoven fractal subsets which exhibit 
power-law spatial scaling behavior with a range of scaling exponents 
(concentration, or singularity, strengths) and dimensions.  The existence of 
MFS implies an underlying multiplicative (or hierarchical, or cascade) 
process. Panoramic column density images of several nearby star-forming cloud
complexes, constructed from IRAS data, are shown to exhibit such multifractal 
scaling, which we interpret as indirect but quantitative evidence for nested 
hierarchical structure.  The relation between the dimensions of the
subsets  and their concentration strengths (the ``multifractal spectrum'') 
appears to satisfactorily order the observed regions in terms of the 
mixture of geometries present, from strong point-like concentrations, to line-like 
filaments or fronts, to space-filling diffuse 
structures.  This multifractal 
spectrum is a global property of the regions studied, and does not rely on 
any operational definition of ``clouds.''  The range of forms of the 
multifractal spectrum among the regions studied implies that the column 
density structures do not form a universality class, in contrast to 
indications for velocity and passive scalar fields in incompressible 
turbulence, providing another indication that the physics of highly 
compressible interstellar gas dynamics differs fundamentally from 
incompressible turbulence.  There is no correlation between the geometrical 
properties of the regions studied and their level of internal star formation 
activity, a result that is also apparent from visual inspection.  We discuss
the viability of the multifractal  spectrum as a measure of the structural ``complexity'' of the
regions studied, and emphasize the problematic dependence of all structural descriptors
on the subjective pre-selection of the region to be described.  A comparison of IRAS 100 $\mu$m column density (not intensity) images
with $^{13}$CO, visual extinction, and C$^{18}$O data suggests that structural details are
captured by IRAS up to at least 30 magnitudes of visual extinction, except in the
vicinity of embedded stars, and that lower-column density connective structure not seen by
other methods is revealed.
\end{abstract}


\section{Introduction}
	Multifractal scaling (MFS) is a property that refers to measures 
or fields for which the local intensity (e.g. density, column density, 
radial velocity, temperature), integrated out to distance $r$ from a given 
position, scales as a power law $r^\alpha$, with the scaling exponent $\alpha$ 
varying from position to position.  If the collection of positions with 
scaling exponents in $(\alpha, \alpha+{\rm d}\alpha)$ can be described as 
a fractal with (Hausdorff) dimension \fa, for a range of $\alpha$, then the 
structure is multifractal, and the function \fa is called the multifractal 
spectrum (MFS; we use MFS for both ``multifractal scaling'' and 
``multifractal spectrum'' when no ambiguity can arise).  This definition 
is not unique in the literature, but is sufficient for our purposes
(see Feder 1988, McCawley 1990, Muzzio \etal 1992).  
The exponent $\alpha$ is called the ``singularity strength'' in the 
multifractal 
literature, since the power law in principle gives a density singularity as $r$ 
approaches zero; here we prefer to use the more descriptive term 
``concentration strength'' for this exponent.  Similarly, the multifractal 
spectrum is often referred to as the ``spectrum of singularities'' in 
the physics literature, but we avoid this terminology here; however, 
since we do use the potentially misleading term ``multifractal 
{\em spectrum}'', 
it is important to remember that ``$f$'' is a dimension of a subset, not a 
filling factor.  

The MFS basically contains information on how the geometry, 
as measured by the dimension, depends on the concentration strength $\alpha$; 
loosely speaking, the \fa curve shows the manner in which regions of 
different concentration strength fill space.  For example, for subregions 
with a given $\alpha$, the dimension will be different for point-like, 
line (or filament)-like, or space-filling structures.  So the 
MFS indicates the distribution of geometries present in a complex structure.  
For a homogeneous fractal the \fa curve collapses to a point; otherwise 
the measure (intensity image) is multifractal, since regions of different 
concentration strength have different dimensions.  A common \fa curve 
between physical processes and/or mathematical models can be interpreted 
as evidence that they belong to the same ``universality class'',\footnote[1]
{
For example, in turbulence the MFS is believed to be ``universal''
at the onset of chaos, because it agrees with the MFS of the 1-dimensional
circle map.
}
at least 
with respect to the \fa function.  However, it is well-known that the \fa 
descriptor (and all other structural descriptors of which we are aware) 
are degenerate with respect to the underlying physics, in the sense that 
systems with very different underlying dynamics can exhibit identical \fa 
curves within the measurement uncertainties; see Bulajich and 
Perez-Pascual (1991), also Chhabra, Jensen, and Sreenivasan (1989). 
We give a simple example in Appendix B.

Multifractal scaling usually occurs in systems that arise from 
multiplicative processes, or ``cascades,'' in which the intensity 
(e.g. density, etc.) at a given position is due to successive 
multiplications along a hierarchical tree (see below), so the existence 
of MFS (and its properties) can be used to infer hierarchical 
(nested) structure, and to constrain the types of physical processes 
that give rise to such a structure.  
The present work attempts to test for MFS in column density maps of nearby 
cloud complexes and to find whether the resulting \fa  function can be used 
to describe the mixture of geometries present in these structures.

Densely sampled extinction, molecular line, H I line, and IRAS maps of
nearby cloud complexes have shown that the gas and dust are organized into
complicated structures, with irregularities present over a wide range of 
scales,
from roughly 0.02 parsecs (resolution limit for the nearest cloud complexes) to
at least a few hundred parsecs (see reviews in Scalo 1985, 1990, Wilson
and Walmsley 1989, Falgarone and Phillips 1991,
Stutzki \etal 1991, Genzel 1991, Falgarone and Perault 1992, Elmegreen 1992,
and references therein).  The geometry and scaling of
this structure may provide a signature of the physical processes that are
involved in star formation, and may be related to important quantities such as
the star formation efficiency and the initial mass function, as suggested by
Henriksen (1991) and Larson (1992).  Elmegreen and Falgarone (1996) and 
Elmegreen (1997) have proposed that the fractal scaling properties of the 
interstellar medium (ISM) are consistent with many of its observed features,
including the 
mass-size scaling, the cloud mass spectrum, and the existence and properties 
of the intercloud medium.  Scaling behavior is the basis for physical 
interpretation in a huge number of applications within physics, with phase 
transitions and turbulence being most prominent.

Since there is no unique way to characterize structure, however, there are many
ways to define structural scaling.  One such method is the perimeter-area
relation (see Feder 1988, chap. 12).  If contours of equal intensity (or
column density, or temperature, etc.) exhibit a power law perimeter-area
relation with a non-integer exponent over a range of scales, this exponent may
be interpreted in terms of a fractal dimension which characterizes the 
manner in
which these curves fill space.  A number of studies have demonstrated such
behavior in local interstellar cloud structures structures\footnote[2]
{
We use the term ``cloud'' loosely to designate an under-resolved 
observed region of space that appears to contain a coherent density 
enhancement, recognizing that under sufficient resolution this apparent 
entity will likely lose its coherence and break up into complex substructure.
}
using extinction 
maps (Beech 1987, Hetem and L\'epine 1993), H I maps (Wakker 1990), 100 $\mu$m 
dust intensity or
column density maps (Bazell and Desert 1988, Scalo 1990, Dickman, Horvath, and
Margulis 1990, Vogelaar, Wakker, and Schwarz 1991) and CO emission maps
(Falgarone, Phillips, and Walker 1991); see also Vogelaar and Wakker (1994).  
Surprisingly, these studies find a
perimeter-area dimension of about 1.3--1.5 for most clouds studied,
similar to the value found for terrestrial cloud and rain areas (Lovejoy 1982)
and slices of laboratory turbulence (see Sreenivasan 1991 for a review).  Even
the HI gas in the M81 group of galaxies appears to exhibit a very similar
perimeter-area dimension (Westpfahl et al 1999). This
result suggests a sort of universality for cloud fractal geometry {\it which is
independent of the presence of young stars or the importance of self-gravity},
and may be related to turbulence (by which we mean phenomena associated with
nonlinear advection), as emphasized by Scalo (1990) and Falgarone \etal
(1991), or some other process, since the dimension of a fractal is extremely 
degenerate with respect to underlying processes and morphological appearance
(i.e. many very different processes and forms can give rise to the same 
dimension).

In the present work we are concerned with a different type of scaling 
behavior, and it is important to appreciate the difference.  If the 
intensity (or column density, or radial velocity, or any measured
quantity) image is represented as a surface whose local height is the intensity
(``mountain range,'' or ``topographic,'' representation), then the dimension
obtained from the perimeter-area relation characterizes the irregularity of
contours formed by horizontal cuts through the ``mountain range.''  However, 
this
dimension does not uniquely or completely characterize the scaling of
the intensity surface itself.  Adding a constant or multiplying by any 
function of the
image intensity will not affect the power-law slope of the perimeter-area
relation, although, for example, the ``spikiness'' of the mountain range may be
greatly affected.  Furthermore, it is commonly stated that if the 
perimeter-area
relation is a power law with exponent $d$, then the clouds have a fractal
dimension $D = 2d$.  However, this gives the incorrect impression that if this
scaling holds then the cloud complex is a statistically self-similar fractal,
when in fact perimeter-area scaling only demonstrates that the ``mountain
range'' is a self-affine fractal.\footnote[3]
{
Self-similarity refers to structures for which each sub-structure
can be obtained from the whole structure by a linear contraction which
reduces the original structure by the same scale factor in all coordinates.
 Self-affine refers to structures whose substructure can be obtained from
the original by a linear transformation in which different coordinates
(e.g. column density and spatial scale) are contracted by different
factors.
}
The structure of the intensity surface of
such self-affine fractals may possess a continuous range of scaling exponents
or dimensions.  

In the multifractal approach used here, the column density 
``surface'' is not sliced at different intensities.  
Instead the present technique is equivalent to partitioning the image 
into different subsets according to the value of the local
concentration strength  
$\alpha$, defined by the exponent of a local power law scaling of the intensity 
integrated out to distance $r$ from a position, if such a power-law scaling 
exists.  The dimension of this subset is not estimated by treating it as a 
collection of curves (contour map) and characterizing the irregularity of 
each curve; the dimension is estimated for the entire subset, treated as a 
collection of points, and includes all the points, not just those comprising 
disjoint closed curves.  No reduction to ``contours'' is involved in the 
process.
As will be seen, the dimensions estimated in this way are very 
different from the perimeter-area dimension, although the relation between 
the two remains unclear to us.  For example, we will show that the column 
density surface 
in the Taurus complex (and the other regions we examined) is a
multifractal, even though it is known to possess a well-defined perimeter-area
relation with a single fractal dimension for all intensities 
(Scalo 1990, Falgarone \etal 1991).  Still, we think that the perimeter-area 
dimension is a useful and illuminating tool that is complementary to other
descriptors.  A detailed discussion of the perimeter-area dimension
applied to the interstellar medium is given in Dickman et al. (1990), to which 
the reader is referred for more discussion.  (Note, however, that the latter
study was based on IRAS intensity images, not column densities as used in the
present MFS investigation.)

It is also important to distinguish the structural scaling examined here 
from the
type of power law correlations which have often been discussed for interstellar
clouds, e.g. cloud size or mass spectrum, or correlations between density or
velocity dispersion and region size.  First, these latter correlations are 
between
properties defined for separate entities, such as ``clumps within clouds'' or
certain operationally-defined types of ``clouds'' located in different
star-forming regions or different locations in the galaxy.  In contrast, the
scaling examined here is a property of a given region as a whole, not a 
correlation between attributes of any operational entities (``clouds''); 
the concept of ``cloud'' does not enter the scaling examined here.  
Second, the
specific form of the quantities whose scaling is being examined here (see 
Section
III below) is directly related to the geometry of the structure, in the sense
that they describe how the structure fills space.

The manner in which the discovery of multifractal scaling
might be related to physical processes is indicated by similar 
investigations in other fields.  
Fully-developed incompressible turbulence in a
variety of experimental situations exhibits an apparently universal 
multifractal
singularity spectrum, whether determined for the spatial distribution of the
dissipation field (see Sreenivasan 1991, 1996 and references therein) or 
for the
velocity signal (Muzy, Bacry, and Arneodo 1991). Thus the \fa
spectrum may reveal an {\it invariant} signature of the underlying nonlinear
fluid dynamical processes in incompressible turbulence (e.g. vorticity tilting
and stretching in three dimensions); i.e. \fa may reflect the essential 
physical processes independently of the type of flow or boundary conditions.  
However, the process reflected by the MFS may be much more generic.  
For example, 
Meneveau and Sreenivasan (1989), Chhabra and Sreenivasan (1992) and others 
have argued that
turbulence is in the same MFS universality
class as certain simple multiplicative processes, providing a guide for
theoretical modeling.  A comparison with direct numerical simulations was given
in Hosokawa and Yamamoto (1990).  Here we show that such a universal \fa 
function does not exist for the density structures of the highly compressible 
cool interstellar medium, even though the structures are multifractal.  

In an 
extragalactic context, Martinez \etal (1990) suggested that the 
three-dimensional distribution of galaxies in the CfA survey possesses 
multifractal scaling, and showed that the corresponding \fa
spectrum could be used to distinguish the Soneira-Peebles hierarchical model (a
homogeneous fractal) from the Voronoi tessellation model (a multifractal), even
though the correlation function for these two models are very similar.  A
study of multifractal scaling in cold dark matter simulations, with an
emphasis on finite size effects, has been given by 
Colombi \etal (1992).  
A number of more recent discussions of the relation between possible 
multifractal scaling and large-scale galaxy clustering have appeared; see 
Sylos Labini and Pietronero (1996) and references given there.  
Multifractal scaling has been discussed in other astrophysical contexts, 
such as the attempt to relate the multifractal spectrum to the geometry 
and underlying physics of the solar photospheric magnetic field by 
Lawrence, Rumaikin, and Cadavid (1993).  Outside of
astrophysics, the use  of the multifractal spectrum as a diagnostic tool is widespread. 
Examples  include various aggregation and coalescence models (see Nagatani 1992), 
characterization of stages of cancer development through analysis of  normal and malignant
tissues (Muller \etal 1991), and analysis of protein  folding through profiles of solvent
accessibilities of amino acid  chains (Balafas and Dewey 1995). 

Multifractal scaling is closely associated with multiplicative processes 
(see sec. II below), such as stretching and folding of fluid elements 
(e.g. Muzzio et al. 1992), and so its existence is sometimes taken as evidence 
for hierarchical structure (e.g. Balafas and Dewey 1995).\footnote[4]
{
In contrast, the existence of other types of power-law scaling 
does not obviously provide evidence for nested or multiplicative structure.  
For example, if a collection of disks with a power law distribution of 
sizes is distributed at random on a plane, the distribution is a homogeneous 
fractal, even though there is no nesting; i.e. the distribution will 
appear (statistically) self-similar as one ``zooms in'' on it, but the 
structure is not hierarchical with respect to nesting.  There are many 
ways to obtain a power law size distribution that do not involve any 
multiplicative processes.
}
Since the hierarchical nature of interstellar cloud structure is not yet 
adequately established or characterized quantitatively (e.g. Houlahan and 
Scalo 1992), a search for multifractal scaling in ISM data is of obvious 
interest.  In particular, Meneveau and Chhabra (1990) and O'Brian and 
Chhabra (1992) have suggested an empirical test for hierarchical processes 
based on the MFS, which we apply below.

In addition, the MFS may be useful in characterizing complex interstellar 
structures in terms of the distribution of geometries, since it describes 
the dimensionality of the subsets as a function of concentration strength.  
As far as we know, other global descriptors of structure (e.g. the 
correlation function) are insensitive to geometry, except perhaps for fairly 
simple structures, like periodically-spaced density fluctuations along a 
filament.  Descriptors can be found to identify local specific forms 
(e.g. minimal  spanning tree for finding filaments, Bhavsar and Ling 1988; 
line and plane  shape detectors based on moments, Luo, Vishniac, and Martel 
1996), but these descriptors do not characterize the global geometry in 
the manner of the MFS.

\section{Data}
We obtained 60 and 100 $\mu$m IRAS images of four different regions of low-mass
star formation from the Infrared Processing and Analysis Center (IPAC).  The
emission seen at these wavelengths is the radiation of warm dust
grains heated by local stars and the galactic radiation field.  IPAC estimated
and subtracted the local zodiacal background, and we removed the galactic
background using a cosecant model profile.  Other background models, such as
fitting a plane or polynomial to local minima, yielded very similar results.
A detailed discussion of these and other points can be found in Verter and Rickard (1998).

The far-infrared emission depends on both the dust temperature and the column
density.  These two quantities may be separated given the 60 and 100 micron
fluxes if it is assumed that all of the dust along a given line of sight is at
the same temperature, the dust is optically thin, and that a power-law dust
emissivity holds.  In such a model, the infrared flux is given by
$F_\lambda=B_\lambda(T_{dust})(1-e^{-\tau})$
where $\tau$ is the dust optical depth and $B_\lambda(T_{dust})$ is the
Planck function.  The dust optical depth was assumed to vary with
wavelength as $\lambda^n$, with n = -1 or -2. This relation may
be iteratively solved for the temperature and column density for every point in
the image.

This method for estimating dust temperatures and column densities assumes that
the dust along a given line of sight is at a constant temperature, which
biases the column density estimates toward the column density of the warmer 
dust.  For embedded sources this warm dust will be concentrated very near 
the source (compared to the line of sight depth), constituting a small 
fraction of the true column density along the line of sight, so the column 
density will be underestimated, sometimes by a large factor.  As a result, 
embedded protostars and H II regions may appear as holes in
the final images.  (These artifactual holes may actually provide a useful 
method for identifying embedded sources, since they are visually very 
distinct.)  For this
reason we mostly chose  regions of relatively low mass star formation for this study, 
since there the stars have minimal effect on the column density images.  A 
more sophisticated technique that provides an estimate of temperature 
variations along the line of sight has been proposed by Xie \etal (1991).

The resulting maps do, however, strongly resemble column
density maps based on extinction and molecular line data away from embedded
stars (e.g. Langer \etal 1989 for the B5 cloud, Snell \etal 1989 for Heiles
Cloud 2 and B18 in Taurus, Jarrett, Dickman, and Herbst 1990 for $\rho$ Oph; 
below and Appendix A for various subregions in Taurus). We emphasize that this
agreement applies to column density maps, and that the intensity maps
can appear significantly different; some previous comparisons of IRAS maps
with molecular line (Abergel \etal 1995) or extinction (Arce \& Goodman 1999) 
images have used 60 and 100 $\mu$m {\it intensity} images.  The reliability of the
column density maps presented here can best be seen by visual comparison with the
recent adaptive grid wavelet transform extinction maps for the same regions
given by Cambresy (1999).  

Four nearby ($\sim$100--200 pc) cloud complexes containing
low- to intermediate-mass young stellar objects were chosen for study:  Taurus,
R Corona Australis, Chamaeleon-Musca, and Scorpius-Ophiuchus.  The last three
regions were divided into 2, 4 and 6 subregions on the basis of their
differing visual appearance and star formation properties.  Table 1 lists the
positions, estimated distances, number of resolution
elements (assuming a resolution of 3 arcmin), and rough estimates of the
projected density of young stellar objects, when available, in each
subregion.  Molecular line maps of these regions with the NANTEN 
telescope are summarized in Kawamura \etal (1999) and Yonekura \etal (2000); 
comparisons with earlier work and additional references can be found there.
Extinction maps that
cover roughly the same areas as examined here can be found in Cernicharo (1991, Taurus), Rossano
(1978a, R Cr A), Gregorio-Hetem \etal\ (1988, Cham-Musca), Rossano (1978b, Sco-Oph), Arce \&
Goodman (1999, Taurus), and  especially Cambresy (1999, Taurus, Sco-Oph, Cham-Musca, and R Cr
A).   The star
formation activity varies from no detected activity in Cham III to widely
distributed low-mass stars (and probably higher-mass B stars---see Walter and
Boyd 1991) in Taurus, to dense and probably bound clusters in the
$\rho$ Oph core and in the R Cr A core.  A discussion of the young stellar 
populations in parts of these
regions with references to the original papers can be found in the review by
Zinnecker {\it et al.} (1993) and the study by Chen \etal (1997).  
The gas in the entire Sco-Oph region has been
affected by interaction with massive stars (see DeGeus 1988).  Figures 1, 2, and 3
show logarithmically scaled column density maps of the
Cham-Musca, RCrA, and Sco-Oph regions. An image of Taurus enhanced by unsharp
masking is shown in Appendix A.  The original Taurus column density
image is given in Scalo (1990)  and a contour map and a ``landscape'' representation of the
region are in Houlahan and Scalo  (1992).\\
~
\begin{center}
~
\psfig{figure=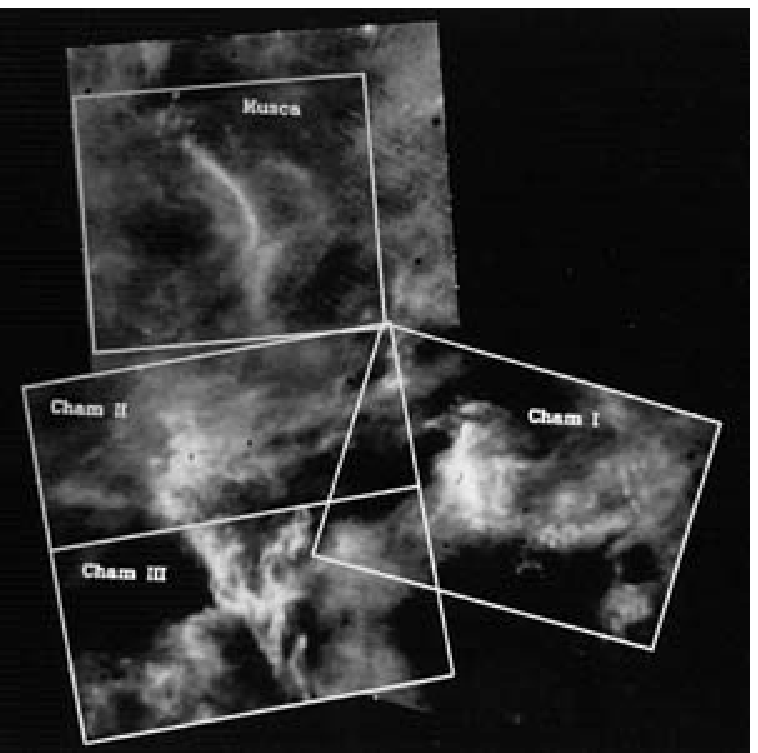}
~
\end{center}
\figcaption[fig1.smaller.eps]{Logarithmically scaled image of the column density
distribution Chamaeleon-Musca.  The small dark holes are embedded protostars
which locally heat the dust, yielding an underestimate of the actual column
density.  The image is a mosaic of three frames, each of which covers
$8\times8^\circ$ on the sky, corresponding to $20\times20$pc 
(assuming a distance
of 140 pc), and contains roughly $6.7\times10^4$ resolution elements.  North is
up and west is to the right.}

\begin{center}
~
\psfig{figure=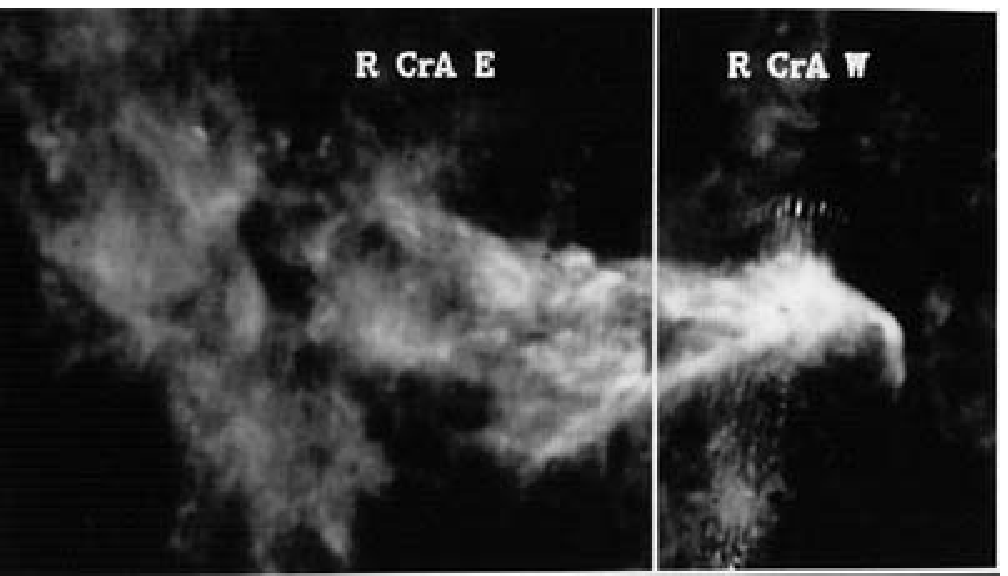}
~
\end{center}
\figcaption[fig2.eps]{Logarithmically scaled image of the column density
distribution around R Corona Australis.  The image covers $14\times8^\circ$ on
the sky corresponding to 32$\times$18 pc (assuming a distance of 130 pc) and
contains roughly 4.5$\times10^4$ resolution elements.  Residual striping
artifacts run vertically through the R CrA W subregion.  North is up and west
is to the right. \label{fig2}}

\begin{center}
~
\psfig{figure=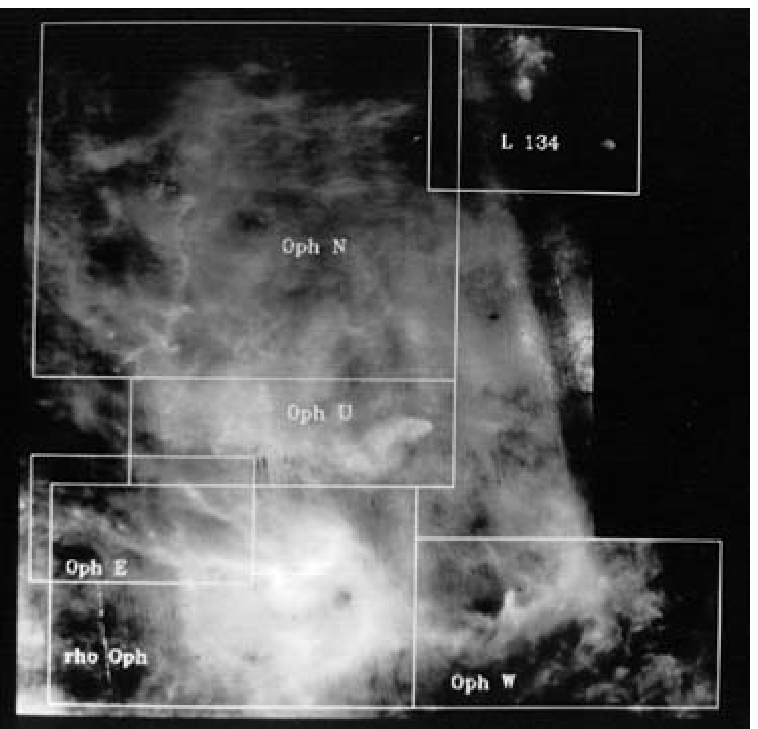}
~
\end{center}
\figcaption[fig3.smaller.eps]{Logarithmically scaled image of the column density
distribution in the Scorpius-Ophiuchus region.  The image covers roughly
$28\times28^\circ$ on the sky corresponding to 61$\times$61 pc (assuming a
distance of 125 pc), and contains 2.5$\times10^6$ resolution elements.  The
light and dark bands running below and to the right of L134 and east of $\rho$
Oph are residual striping artifacts.  North is up and west is to the right. \label{fig3}}

There is considerable uncertainty concerning the degree to which IRAS 100 $\mu$m
column density N$_{100}$ represents the true column density. A thorough
discussion, with application to cirrus clouds, has been given by Verter and
Rickard (1998). Generally the  relation between N$_{100}$ and
$^{13}$CO column density exhibits a great deal of scatter, yet the resulting spatal images
of N$_{100}$ do closely resemble the structure of $^{13}$CO and extinction maps (e.g.
Langer
\etal 1989 for the B5 cloud, Snell \etal 1989 for Heiles Cloud 2 and B18 in Taurus,
Jarret, Dickman, and Herbst 1990 for $\rho$ Oph). Much of the scatter probably arises from
a  combination of the residual striping, the embedded protostar effect discussed above,
and perhaps the background subtraction at the lowest intensity levels.  (We point out 
that the large scatter found in studies like Abergel \etal 1995 is most likely due to  the
use of IRAS intensities rather than column densities.)  However none of the effects 
seriously compromises the overall spatial structure except in  very localized ways,  or at
the lowest intensity levels.  The latter effect forces us to consider only  positive
moments in the MFS (i.e. the ``hole'' geometry is not considered -- see  sec. 4.1 below). 

Since the MFS responds to the  global structure on all scales,  it is the similarity
between the spatial structure in N$_{100}$ and other tracers  that is most important. We
have compared the images of N$_{100}$ with available 
$^{13}$CO and extinction maps of five subregions in Taurus, and the results 
show excellent agreement.  An example, the  subregion L1529 filamentary substructure 
in Taurus is shown in Appendix A. The only serious disagreement occurs for the very 
opaque L1495 subregion. However we find that the N$_{100}$ image for this subregion 
does agree much better with the C$^{18}$O  image of Duvert, Cernicharo, and Baudry (1986). 
This suggests that N$_{100}$ is a better tracer of column density than either 
$^{13}$CO or visual extinction, both of which have severe optical depth effects that 
set in around A$_{\rm V} \approx 3-5$. The case of L1495 suggests that the upper
limit on A$_{\rm V}$  for the validity of N$_{100}$ may be around 30 mag, rather than the
10 mag limit  suggested by Jarrett \etal (1989) for the 60 $\mu$m column density. We also
find  surprisingly good agreement between the $\rho$ Oph core N$_{100}$ image derived here
and the submillimeter image given by Motte \etal (1998), considering the much better 
resolution of the submillimeter image, suggesting that the A$_{\rm V}$ limit for N$_{100}$ 
(except along lines of sight to embedded protostars) may exceed 50 mag.  Thus the IRAS column
density upper limit may be similar to that of infrared extinction (e.g. Lada, Alves, and
Lada 1999), except that IRAS is locally thwarted by embedded stars, while infrared 
extinction requires smoothing to reduce noise introduced by structure in the background star
positions.  

A strong indication of the validity of the
IRAS column density images is the excellent visual agreement of our images with the adaptive
grid/wavelet decomposition optical extinction maps of several of the regions by Cambresy
(1999).  Another favorable example is that the "cores" mapped in the C$^{18}$O study of a large
part of Ophiuchus by Tachihara, Mizuno, \& Fukui (1999) can be seen to correspond to the
highest-column density peaks in the IRAS images of Ophiuchus displayed in Fig.3; the IRAS column
density images indicate that the apparently isolated $C^{18}O$ cores studied by Tachihara et
al., especially in the regions we refer to as Oph N and U, are actually connected in a web of
elongated or filamentary forms whose column density was too small for the $C^{18}O$ survey.  

Overall, these comparisons lead us to be optimistic about the validity of the  IRAS column
density maps, except near embedded sources or at very low column densities  (which we do not
use in the subsequent analysis). As remarked above, we found our results  to be robust with
respect to the adopted power law form of the far-infrared emissivity. Lastly, we found no
problems associated with producing seamless mosaics of the IRAS maps, except that it was
technically laborious.

We emphasize that a major advantage of the IRAS data is that it can probe a very large
range of column densities compared to other methods, roughly from about 
$2\times 10^{21}$cm$^{-2}$ (see Weiland eta l 1986, Verter and Rickard 1998) to 
$6\times 10^{25}$cm$^{-2}$ (assuming that the ratio of A$_{\rm V}$ to $\tau_{100}$ is
a few thousand).

\section{Calculation of Multifractal Statistics}
The multifractal spectrum is a method of characterizing images
and point sets which possess a type of statistical scale invariance called
multifractal scaling.  The local density profile around a
point in the image is characterized by a scaling exponent called the
singularity strength (here we use the term ``concentration strength'').  
If $P_i(L)$ is the integral of the image intensity
(column density in the present work) within an aperture of radius $L$ centered
at position $i$, then the concentration strength $\alpha$ is defined by
$P_i(L)\propto L^\alpha$.  For a two-dimensional homogeneous image $\alpha = 2$,
while for a sharp spike $\alpha = 0$.  Thus, the concentration strength measures
the local ``spikiness'' in the image.  The function \fa  measures how all
the points in the image which have a given concentration strength are
distributed and it is the Hausdorff dimension of that point set.  The
numerical calculation of \fa has proven to be delicate.  Difficulties with 
procedures used in multifractal analysis have been discussed by Chhabra, Jensen, \& Sreenivasan
(1989), Yamaguti and Prado (1995) and Veneziano, Moglin, and Bras (1995).  The latter 
paper, in particular, emphasizes that the \fa spectrum only generates the 
upper envelope of the true MFS, while missing interior points not in the upper 
envelope, and generates spurious points if the actual MFS is discrete (a 
bifractal would be an extreme example). We selected
the canonical method  of Chhabra and Jensen (1989), which is computationally
efficient, avoids some of the problems faced by other techniques, and
provides an intriguing mathematical connection to methods used in statistical
mechanics.  (See McCauley 1990 for a discussion of ``thermodynamic 
formalism'' for multifractals.)

The canonical method for computing \fa begins by partitioning the image
into boxes of width L.  If $P_i(L)$ is the intensity integrated over box $i$,
then \fa is given implicitly by
\beq
f(q)=\displaystyle\lim_{L\rightarrow0}~{{\sum\nolimits_i}\mu_i(q)\thinspace
{\rm log}\thinspace\mu_i(q)\over {\rm log}~L}
\eeq
\beq
\alpha(q)={\sum\limits_i}\mu_i(q)\alpha_i=\displaystyle\lim_{L\rightarrow0}~{{\sum\nolimits_i}\mu_i(q)\thinspace
{\rm log}\thinspace P_i\over {\rm log}~L}
\eeq
where $\mu_i(q)=P^q_i(L)/{\sum\limits_i}
P^q_i$ are the $q^{th}$ coarse-grained moments of the original image and the
sums cover the entire image.  For real data sets, the limits are estimated by
least square fitting of ${\sum\limits_i}\mu_i\thinspace {\rm
log}\thinspace\mu_i$ and $\sum\limits_i\mu_i\thinspace {\rm log}\thinspace P_i$
as a function of ${\rm log}\thinspace L$ for different $q$ values.  If the 
linear
fits are good and the slope varies with $q$, then the data set is said to 
possess
multifractal scaling over the specified range of scales.  If $f$ and $\alpha$
take on only a single value for all moments $q$, then the image is a 
homogeneous
fractal.

The order parameter $q$ can be shown to be the slope of the \fa curve, so the
\fa curve for multifractals is a convex downward opening curve peaking at the
point where $q=0$.  The value of $f$ at that point is the Hausdorff dimension of
the support of the measure, which is 2 for the images considered here, while
$\alpha(q=0)$ is just the average concentration strength of the image.  At the
point where $q=1$, the \fa curve must lie on the diagonal line \fa=$\alpha$. 
The point at $q=\infty$ characterizes the regions in the image which have the
smallest values of $\alpha$ which may be thought of as the sharpest spikes. 
Similarly, the $q=-\infty$ point characterizes the regions which have the
steepest holes, i.e. with the largest values of $\alpha$.  The \fa function is
related to the so-called generalized or Renyi dimensions $D_q$ (which
characterize the scaling of the qth order partition function 
${\sum\limits_i}P_i^q(L)$) through a Legendre transform.

The dependence of the MFS on the parameters of a simple ``toy model''
multiplicative process is illustrated in Appendix B.

The referee has pointed out that the formalism involved in obtaining the MFS seems rather
abstruse and non-intuitive, and suggests the general principle, with which we agree, that
any empirical descriptor should bear a proportional requirement of success in terms of
physical interpretation in exchange for complexity of effort.  While we agree that the
canonical formalism adopted here is not at all transparent, there are several points which
suggest that the use of the MFS does not violate the above general principle. First, the computational complexity, if measured in cpu time, is actually smaller
for the MFS than for the correlation function.  Second, it must be recognized that the
degree to which a descriptor seems obscure or complex depends inversely on the familiarity
of the user.  For example, the meaning and significance of the correlation function often
seems obscure to those who have not dealt directly with it.  Concerning the ability of the
MFS to reveal some underlying physics, the reader may judge from the rest of the paper.  We
only remark that the MFS does allow us to infer the presence of some multiplicative
process, and also shows a fundamental difference between incompressible and supersonic
turbulence with respect to universality class, as discussed below.  By comparison, studies
of the velocity correlation function, for example, have revealed few significant clues to
underlying physical processes, as far as we know.  

\section{Results}
\subsection{Multifractal Scaling}

Figure 4 shows ${\sum\limits_i}\mu_i\thinspace {\rm log}\mu_i$ and
${\sum\limits_i}\mu_i\thinspace {\rm log} P_i$ as a function of 
${\rm log}~L$ for various
positive $q$ values in the Taurus region.  The range of box sizes used for the
fits was 10 to 100 pixels, which corresponds to linear sizes of 0.4 to 4 pc.  The lower
limit is set by the resolution.  For example, the adopted IRAS angular resolution of 3
arcmin corresponds to a linear resolution of 0.13 pc for Taurus, at the assumed distance
of 140 pc (similar to all the regions studied here).  Image structure is generally
severely compromised for images smaller than 3-4 resolution elements, or 0.3-0.5 pc for
the regions studied here.  The upper limit is set by the fact that at larger scales a
significant fraction of the boxes used to compute the MFS overlap the edges of the images. 
{\it We emphasize that the image size must be significantly larger than the upper limit
for the range examined for scaling behavior; otherwise edge effects distort the scaling}. 
Related edge effects give comparable distortion to  other descriptors of structure (e.g.
correlation function, see Houlahan and  Scalo 1990 for an explicit demonstration).  The
panoramic nature of the IRAS  images make it possible to satisfy this requirement. 
However, errors in calibration and background removal dominate and distort the scaling for
negative $q$ values, which accentuate the low-flux areas.  As a result, the scaling for
negative $q$  values could not be investigated.  The good linear fits (at least for lower
$q$ values), with different slopes for different $q$ values, exhibited in Fig. 4
indicate that the column density structure in Taurus is multifractal, at
least over the range of scales examined.  These fits are typical of nearly all
the subregions.  Fig. 5 shows the scaling behavior of all of the subregions 
for a $q$ value of 8.  We conclude that all the
regions used in this study exhibit multifractal scaling.  
This result demonstrates that there are no preferred scales in the column density
structure over a factor of about ten in scale (roughly 0.4 to 4 pc in the regions
studied).  Such a small range does not provide a very strong constraint on preferred
scales, however.

Multifractal scaling is usually associated with multiplicative 
processes, involving hierarchical nesting or a ``cascade.''  Meneveau 
and Chhabra (1990) and O'Brian and Chhabra (1992) have shown that the 
two-point correlation function of the spatial field of scaling exponents 
$\alpha$, $C_\alpha(s)$, should exhibit a logarithmic decline with 
separation $s$ for random 
isotropic multiplicative processes, providing a test for hierarchical 
structure.  We have applied this test to the Taurus region by 
constructing an images of $\alpha$, and then calculating the correlation 
function 
$C_\alpha(s)$.  The result is shown in Fig. 6, where it is seen that 
a significant range of logarithmic decay of $C_\alpha(s)$ exists.  The fit to a 
logarithmic decay is even better than the turbulence examples shown in 
O'Brian and Chhabra (1992).  
We postpone a discussion of the other regions to a separate 
publication.
~

\begin{center}
~
\psfig{figure=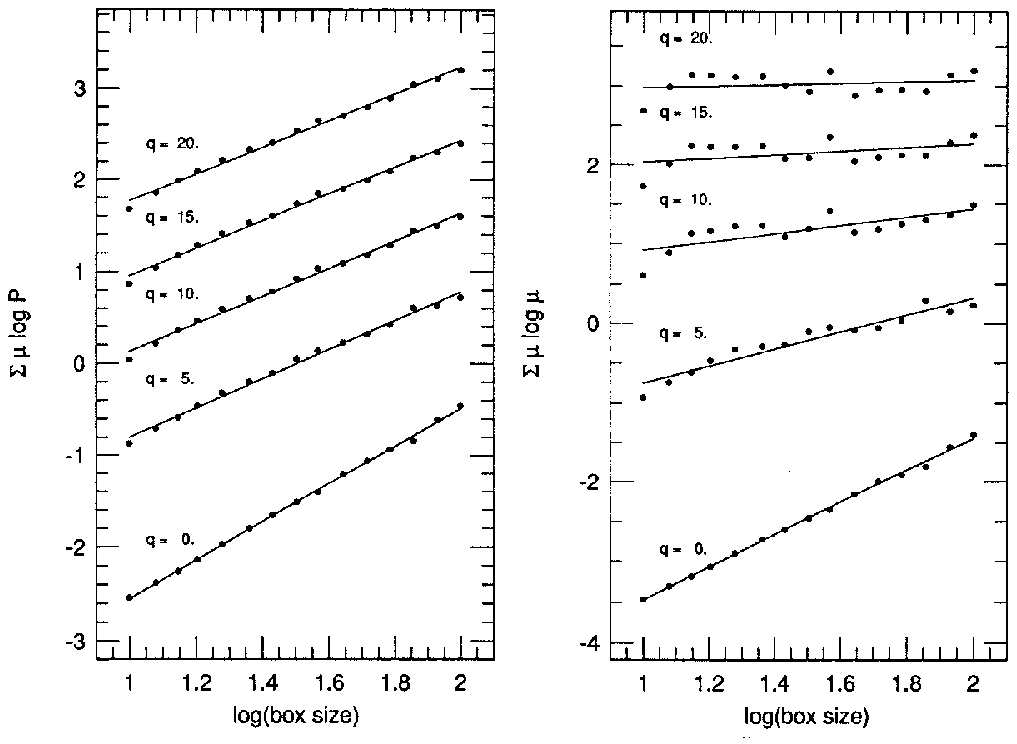}
~
\end{center}
\figcaption[fig4.smaller.eps]{Multifractal scaling in Taurus.  The slopes of the linear regression
fits (indicated by straight lines) in the left panel gives the value of
$\alpha$ for each $q$ value while the slopes in the right panel give the valves
of $f$.  The units of the abscissas are log arcminutes.  At an assumed distance
of 140 pc, the box size limits range from 0.4 to 4.0 pc.  Arbitrary constants
were added to each set of points to prevent the fits from overlapping. \label{fig4}}
~
\begin{center}
~
\psfig{figure=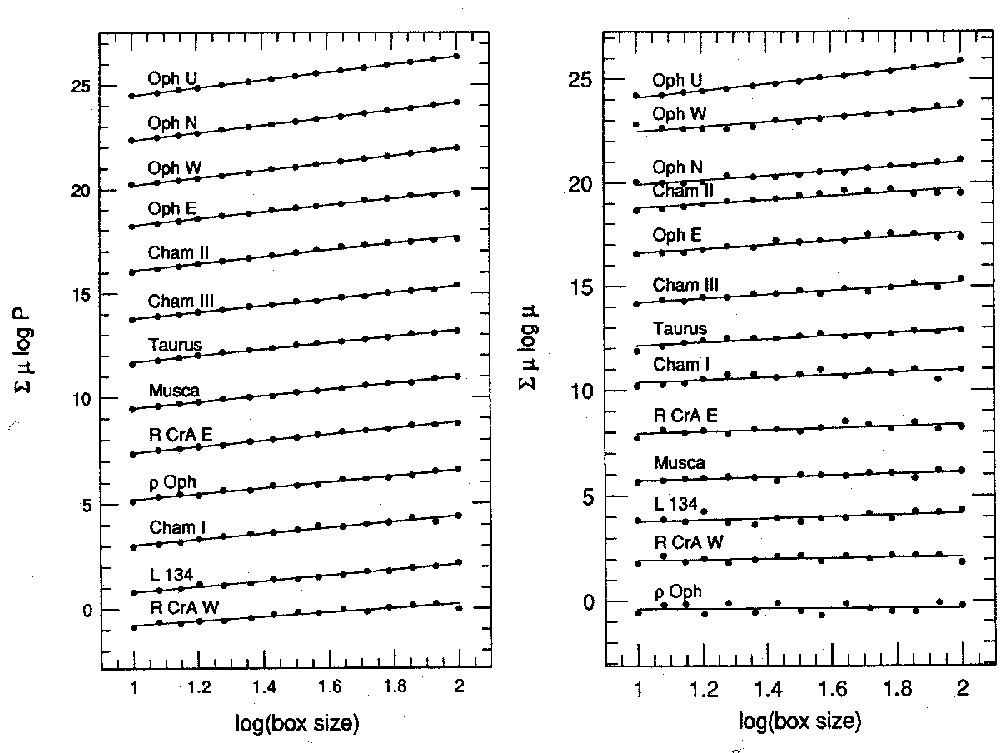}
~
\end{center}
\figcaption[fig5.smaller.eps]{Multifractal scaling for all the regions.  The fits for $\alpha$ and
$f$ at $q=8$ are shown for all the regions.  The fits are ordered by their
slopes, with the steepest slopes at the top.  Each set of data points were
arbitrarily shifted up or down to prevent overlapping on the plot. \label{fig5}}

\begin{center}
~
\psfig{figure=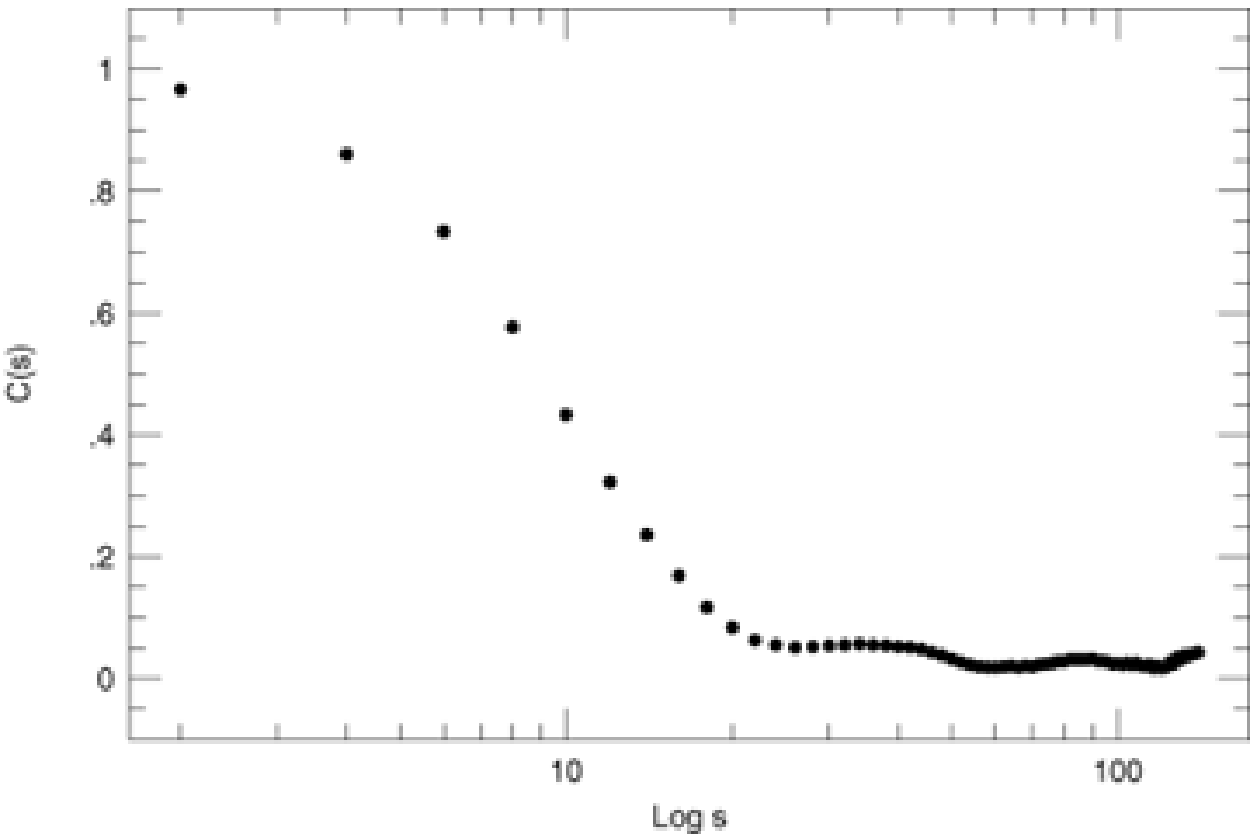,height=3in,width=4in}
~
\end{center}
\figcaption[fig6.smaller.eps]{Correlation function for the image of concentration
or ``singularity'' strengths $\alpha$ for the Taurus region.
The significant range over which logarithmic decay occurs is interpreted
as evidence for nested hierarchical structure (see text). \label{fig6}}

\subsection{Multifractal Spectrum}

Figure 7 shows the \fa curve for positive $q$ values for each of the
subregions.  These \fa curves were found to be fairly insensitive to the choice
of background or to the adopted power law in the variation of dust emissivity
with wavelength. The error bars represent the $\chi^2$ fitting error in
determining $f$ and $\alpha$. The curves are ordered by their width, i.e. by
$\alpha(q=20)$.  The \fa curves for the subregions display a variety of 
shapes. 
 
We have attempted to determine whether the subregions could be uniquely ordered
according to some overall property of the \fa curves.  Most of these attempts
consisted of plotting the positions of the subregions in a two-dimensional
diagram, each of whose axes correspond to some property of the \fa curves.  For
example, one such property is the range of $f$ values covered by a fixed 
range of
$q$ values which we chose as $4<q<20$ (since the $q<4$ points are very similar
for most regions).  This parameter $\Delta f\equiv f(q=4)-f(q=20)$ may be 
thought
of as a ``dimensional diversity,'' since images dominated by a single strong
point-like concentration will have $f$ values close to zero for most 
positive $q$
values (low dimensional diversity), while those with a range of geometries
(from surface-like to point-like) will have a large value of this parameter. 
Figure 8 shows the observed regions in the $\Delta f-\alpha(q=20)$ plane, where
$\alpha(q=20)$ indicates the strength of the strongest concentrations.  A fair
correlation is present, but this is expected because the \fa curve must, by its
definition, remain below the diagonal line $f=\alpha$, so the range in $f$
values must decrease for curves which have smaller $\alpha$ values; i.e. images
which possess strong dominant concentrations will usually have smaller
diversities.  

The ordering of the regions in Fig. 7 does correspond to the visual 
appearance of the 
geometry of the structures in the column density images.  For example, 
R CrA W has the broadest \fa and has the largest ``point-like'' 
contribution, consistent with the fact that the region is dominated by 
an extremely high-column density core.  L134 and Cham I are also 
``point-like,'' although not as concentrated as R CrA W.  At the 
other extreme, the \fa curves for Oph N, W, and U are the narrowest of 
all the regions, in agreement with the smoother, diffuse, appearance and 
lack of strong concentrations seen in the column density images.  The 
vertical extent of the \fa curves for the most diffuse regions (Oph N,W,U) 
may be unreliable because of finite size effects.  For example, it is 
well-known that a white noise field, which should appear as a single point 
at $\alpha = f = 2$, will instead exhibit a very narrow measured \fa that 
extends downward to smaller dimensions because of the finite size of the sample.
Thus the dimensional diversity f(q=4)-f(q=20) of Fig. 8 could be 
significantly smaller for these regions than calculated here.  The 
intermediate regions (Musca, $\rho$ Oph, R CrA E, Cham II, Cham III, Taurus, 
and Oph E) all exhibit some quasi-linear or filamentary structure, mixed 
with point-like concentrations and more diffuse structures.  Oph E does not 
contain any prominent filaments, being primarily diffuse like Oph N,W,U, but 
it does contain an irregular ``ridge'' that may reflect the presence of 
an ionization/shock front, and it is probably this feature that shifts 
the \fa spectrum toward the regions with prominent filaments.  Notice also 
that the total R CrA region visually resembles the $\rho$ Oph region dense 
core with ``streamers'' to one side.  It  is because we have divided 
R CrA into two subregions (the core and the streamers) that the core, 
R CrA W, separates so clearly from the rest of the regions in Fig. 8. 
 
This result illustrates the problematic dependence of all structural descriptors 
on the subjective pre-selection of the region to be described.

\begin{center}
~
\psfig{figure=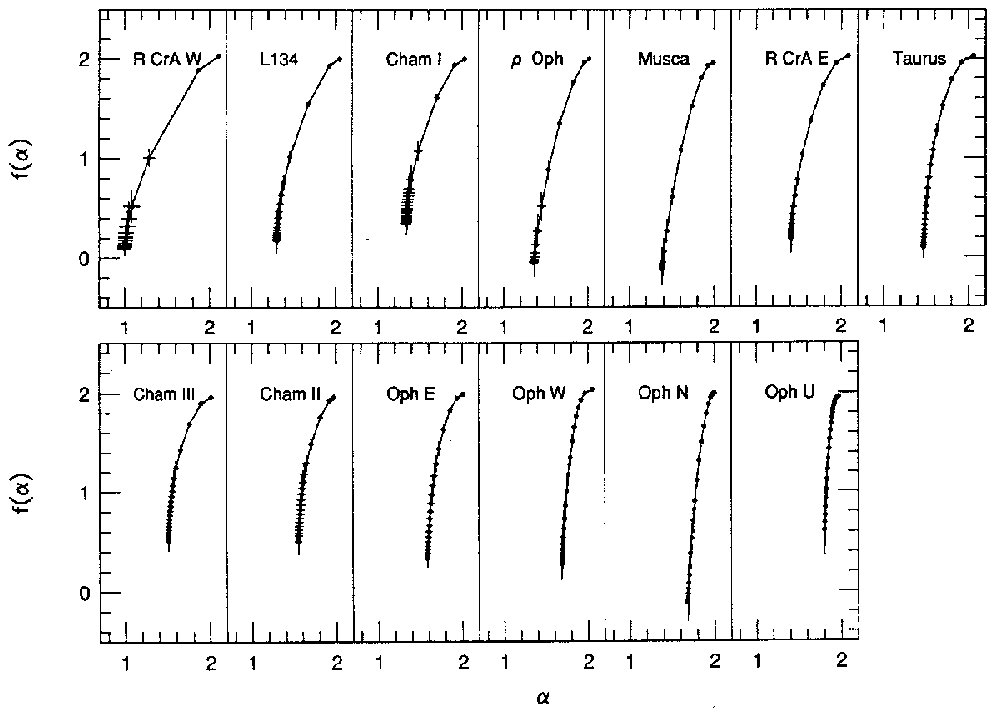}
~
\end{center}
\figcaption[fig7.smaller.eps]{Spectrum of singularities of the column density maps of all the
subregions.  Due to background and calibration uncertainties only the left
positive-$q$ portion of the $f(\alpha)$ curves were calculated.  The regions
are ordered by the strength of the strongest singularities present, i.e. by
$\alpha(q=20)$ which is the width of the  $f(\alpha)$ curve. \label{fig7}}

\begin{center}
~
\psfig{figure=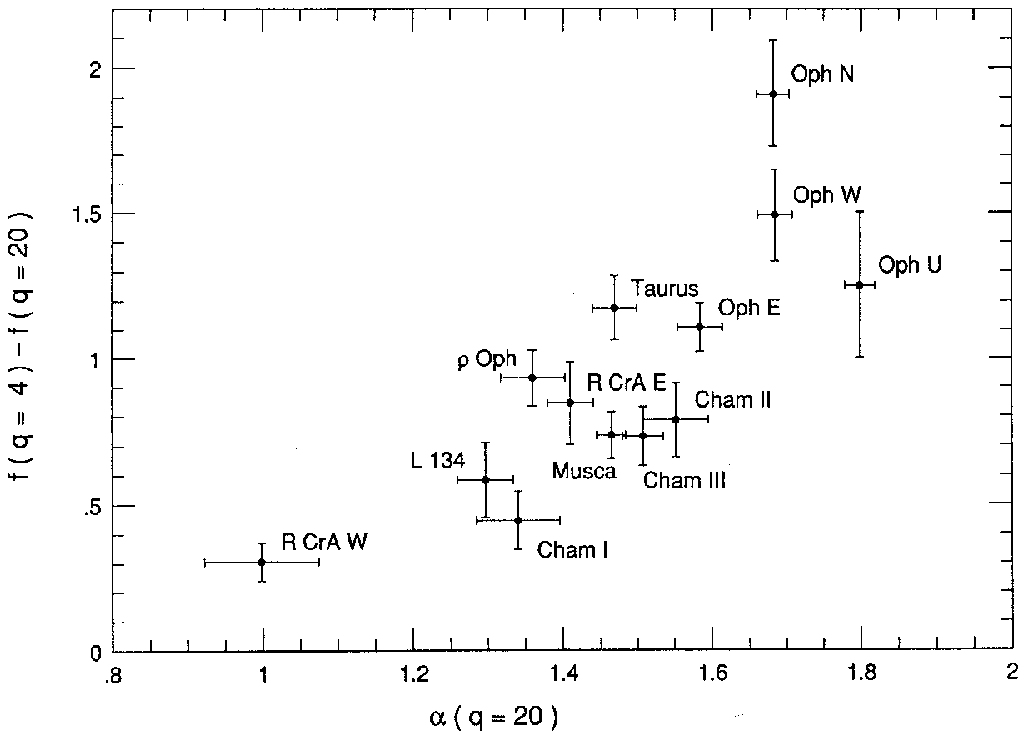,height=4in,width=6in}
~
\end{center}
\figcaption[fig8.smaller.eps]{The ``dimensional diversity,'' taken as $f(q=4)-f(q=20)$, is plotted
as a function of the strength of the strongest singularities present in the
region, i.e. $\alpha(q=20)$.   \label{fig8}}

Since the ordering of regions described above does involve the degree of 
concentration, it might be erroneously thought that the properties of \fa 
at large $q$ are simply reflecting the degree of density contrast.  We have 
examined the probability distributions of column density for all the 
subregions and their moments.  The standard deviation (s.d.) relative to
the mean column density varies from 
0.20 (Musca) to 0.55 (R CrA W).  So R CrA W is extreme in both \fa 
ordering and column density contrast, and, similarly, Oph U and Oph N 
are rather extreme at the other end, ranking 10 and 8 (out of 11) in 
standard deviation, respectively.  However there are exceptions that 
make it clear that the \fa ranking is primarily geometrical, even if 
density contrast does correlate with geometry in some cases.  For example, 
Musca has by far the lowest s.d., yet lies in the middle of the regions in 
Fig. 9 because it is primarily a simple filament.  L 134 is ranked 7 in s.d. 
(0.34), yet it, along with Cham I, is, after R CrA W, the most 
``point-like'' of the regions, not at all like Oph N, which has 
a similar s.d.  Another example is R CrA E, which is ranked 2 in s.d. 
(0.46), even though it is near Musca (rank 11), Cham II (rank 9), and 
Cham III (rank 6) in Fig. 8 due to the prevalence of line-like structure 
in all these regions.  Oph W is among the three most 
``space-filling'' or homogeneous 
regions according to Fig. 8, yet its standard deviation is rank 5 (0.38).  
Clearly the \fa indicator is responding to geometry and not density contrast 
in these cases.  

It should be emphasized that, while some properties of \fa do reflect some 
of the gross character of the geometry of these regions, and hence may be 
useful descriptors for classification and comparison with simulations, \fa 
definitely does not capture other aspects of structural detail.  For example, 
the Cham II and Cham III regions have
very similar \fa curves but different visual appearances: The Cham III region
looks more ``disrupted'' or filamentary (even though it contains few, if any, 
young stellar objects; see below).  Thus, the \fa spectrum, like all other 
structural descriptors that have been proposed, suffers from severe 
compression of structural information.  On the other hand, the utility 
of different structural descriptors, even if, or just because, they are 
highly compressed, is their ability to capture different aspects of the 
structure, and in that regard the geometrical information expressed 
compactly by \fa is important because those aspects are not expressed 
by other descriptors.  However it appears that no quantitative descriptor 
will be capable of capturing any but a small fraction of the aspects of 
structure apparent to the eye/brain.  This is of course a common and 
severe problem in all applications of pattern analysis when description, 
rather than specific recognition or identification, is the goal.

Our method of ordering the regions in terms of some property of the \fa 
curve is not unique, and we considered other approaches.  For example, we 
tried to order the regions according to the distance of \fa from some
fiducial reference \fa curve, defining distance as in Adams (1992) and 
Adams and Wiseman (1994).  For the fiducial curve we tried various colored 
noises, stochastic multiplicative processes, and the mean \fa for the 
observed region.  However it was found that the resulting ordering depended 
strongly on the choice of the standard \fa curve.  We suspect that this 
dependence of ordering on the adopted reference function will be a 
general problem for this approach to classification.  Instead we would 
suggest that the variation of ordering with respect to fiducial reference 
function is itself an important descriptor of complexity, since it is an 
indication of the number of ways in which the collection of regions can be 
viewed; i.e. a measure of the ambiguity presented by the structure.  This 
has been a common theme in recent discussions of the meaning of complexity, 
although we know of no quantitative measure that has been proposed.

The relation of \fa to the stellar content of the different regions is also of interest. 
It is generally thought that the star  formation properties of 
cloud complexes might be related to the geometry of the dust and gas, 
although even the direction of the causation is unclear.  Larson (1992) 
proposed that the stellar initial mass function may be a consequence of 
the geometry of the gas from which stars form.  From a different point of 
view, it is generally thought that the presence of young stars should alter the structure of the gas in the parent cloud 
complex in a severe manner, although no quantitative descriptor of this 
effect has, to our knowledge, been proposed.  For this reason, several star
formation indicators were compiled for each of the regions, including the
surface density of young stellar objects, the ratio of stellar luminosity 
to gas mass, and the star formation efficiency.  (See the references in Table 1; 
also the study of star formation in these regions by Chen \etal 
1997.)  No correlation with the shape of the \fa curve was found.  The 
reason for this lack of correlation is fairly obvious: {\em There is simply 
no one-to-one correspondence between gas morphology and stellar content.}  
For example, while L134, with little internal activity, has a much smaller 
column density than either $\rho$ Oph or R CrA W, it certainly does have 
a ``core-like'' appearance, and so is geometrically similar to the 
cores with strong star formation activity.  When discussing ``cores'', 
this lack of correspondence may seem trivial, since it might simply be due 
to the fact that a region without internal activity hasn't had time to 
begin star formation, or has too small a mass to be bound, etc.  However 
other examples show that the lack of correspondence goes further than 
regions that appear to comply with conceptions of quasi-spherical 
``clouds.''
Both Taurus and Cham III have similarly prominent filamentary structure, 
but Cham III is devoid of star formation activity.  Musca is another 
filament with little internal activity; the same applies to the streamers 
extending away from the $\rho$ Oph and R CrA cores.  The differences may be 
due to the ages of the regions or because the gas morphology reflects 
external forces (particularly for Musca) as well as internal processes 
like self-gravity or hydrodynamic instabilities.  Perhaps the most 
interesting comparison is Cham II and III.  As mentioned above, Cham III 
appears significantly more ``disrupted,'' even though it has essentially 
no measurable internal star formation activity.  These examples suggest that 
it will be quite difficult to understand any relationship between star 
formation and gas structure, given the small number of regions that can 
be mapped at sufficient resolution, and the fact that we observe each 
region at different effective ages.  Still, it would be interesting to 
carry out the type of geometrical analysis presented here for the column 
density structures in regions with strong massive star formation.  The 
IRAS data are unsuitable for such maps because the heating by the massive 
stars probably seriously distorts the derived column densities, as explained 
in sec.II.  The Bell Labs $^{13}$CO survey provides sufficiently dense 
mapping of 
massive star formation regions for estimates of \fa (for the column densities 
and the radial velocities), and should be attempted.  Based on the results 
found here, it is likely that \fa will primarily reflect the mix of strong 
``point-like'' structures (very dense ``cores'') and filamentary 
structures, both of which are known to exist in these regions.  Thus, 
there is no reason to expect the properties of the \fa curve to be  
significantly different from those found for, say, Taurus or $\rho$ Oph.  

\section{Summary}
We have investigated the possibility of multifractal scaling in several regions
and subregions of local cloud complexes using column
density images with a very large spatial dynamic range constructed from IRAS
data.  The major results are as follows.

1.  All the regions and subregions exhibited multifractal scaling over the 
range
0.4 to 4 parsecs, the largest range available for investigating scaling using
the available images and this technique, even though the images covered a much
larger range, because of the sensitivity to finite size and resolution effects 
(see also
Takayasu and Suzuki 1991).  That the column density images do exhibit
multifractal power law scaling is a new and surprising result, which should
provide an important constraint on theoretical models for the physical origin 
of the structure in star-forming clouds.  In particular, since MFS
is commonly associated with multiplicative processes, the existence of
MFS in cloud complexes can be taken as indirect evidence for nested
hierarchical structure.  We have shown that the correlation function
of the $\alpha$-images of Taurus exhibits a distinct range in which
logarithmic decay occurs, consistent with predictions for a hierarchical 
multifractal process.

2. Contrary to results obtained for the dissipation fields and passive
scalar fields of incompressible turbulence (Sreenivasan 1996 and references
therein; also Muzy \etal 1991), the \fa curves for these interstellar regions
have a variety of shapes indicating that they do {\it not} belong to a single
universality class with respect to \fa.  This result suggests to us that
supersonic magneto-gravitational ``turbulence'' does not possess a statistical
equilibrium range, or that any such quasi-equilibrium evolves significantly 
with
time or depends on initial or environmental conditions.  This is in contrast to
the similar perimeter-area dimension found in previous studies of local
star-forming cloud complexes.  However it remains to be seen whether the
radial velocity images of these regions are multifractal, and whether
they exhibit a universal MFS.

3.  The forms of the \fa curves for these regions do seem to correlate fairly
well with the mixture of geometric forms seen visually.  Thus the MFS appears
to be a useful descriptor for capturing this aspect of structure, and
should be useful in combination with other descriptors in eventually
obtaining a quantitative classification scheme for interstellar structure.
However we find no correlation between the forms of the \fa curves and the
level of internal star formation activity. This lack of correlation between 
morphological structure and star formation activity can be
easily seen directly in several examples by eye.

The fact that all the regions {\em do} exhibit well-defined
multifractal scaling provides a new constraint on theoretical models for cloud
evolution and star formation.  Simulations of realistic ISM evolution
are approaching sufficient resolution to allow a search for multifractal
scaling, although no such search has yet been conducted.
It is known that simple models for both hierarchical
fragmentation (e.g. sequential probabilistic partitioning, as has been used to
model the \fa spectrum of incompressible turbulence) and coalescence
(aggregation) can result in multifractal scaling.  However, in the
fragmentation case the dominant physics is unclear, since the structure 
might be
produced by nonlinear fluid advection, as in incompressible turbulence, or by
repeated gravitational instability or other fluid instabilities.  In the case 
of
coalescence, multifractal scaling has only been demonstrated for Monte Carlo
simulations in which spatial correlations were built in using a specific power
law correlation function (Menci \etal 1993).  It is hoped that the present
results will motivate theoretical searches for multifractal scaling in
simulations of cloud evolution based on the hydrodynamic equations.  Although
very large spatial dynamic range is required, even a negative result would allow one to
conclude that an important physical process has been neglected, while a positive result
might allow an identification of the physical processes that are responsible
for the variation in \fa found among different star-forming regions.

We thank the referee, whose comments on an earlier form of this paper led to
considerable re-thinking and improvement in the presentation.  We are
grateful to Mark Heyer for providing the digitized $^{13}$CO data for Taurus,
Chris Brunt for useful correspondence, and Fran Verter for helpful comments
concerning the treatment of the IRAS data. This work was supported by NASA  grant NAG5-3107
and a grant from Texas Advanced Research Projects.

\appendix
\section{Validity of IRAS Column Densities}

In order to understand the degree to which the IRAS 100 $\mu$m column densities, 
N$_{100}$, represent the true column density distribution, we have compared our 
N$_{100}$ images to images of five subregions in Taurus constructed from $^{13}$CO
(Heyer  \etal 1987), visual extinction A$_{\rm V}$ (Cernicharo and Bachillar 1984),
and, in the case of L1495, C$^{18}$O (Duvert \etal 1986).  These subregions are
L1529(=B18), B217, L1495, L1536, and Heiles Cloud 2. Both $^{13}$CO and A$_{\rm V}$ were
available for Heiles Cloud 2 and L1529, while only $^{13}$CO was available for B217 and
only A$_{\rm V}$ for L1506, L1495, and L1536.

A large-scale logarithmically scaled N$_{100}$ image of Taurus is presented in Figure 9. 
The image has been processed by unsharp masking to bring out the smaller-scale structure. 
The unsharp masking is equivalent to artificially modifying the column density power
spectrum by putting more power at larger wavenumbers.  We were careful to verify that this
image enhancement did not introduce any spurious or artifactual features.  The 
calculations in the text and the comparisons described below are based on unprocessed 
images.

The comparisons all yielded surprisingly positive results in the sense that the resulting
structures are very similar in visual appearance.  A comparison of
$^{13}$CO, N$_{100}$, and A$_{\rm V}$ for the L1529 subregion (the filament in the lower center
of Fig. 9) is presented in Figure 10.  The $^{13}$CO and A$_{\rm V}$ images were taken in
digitized form directly from the original data, but have been smoothed to the adopted
resolution of the IRAS data.  The agreement in morphology is good, especially compared to
the $^{13}$CO data.  Notice the nearly circular ring extending from the middle northern
part of this substructure, which can be seen to various degrees in all three maps,
although its ring-like form is barely discernable in the extinction image.  We concur with
Heyer {\etal}'s conclusion that the extinction errors must be considerably larger than
usually quoted in the literature.
	
Fig. 10 is typical of the five comparisons that were carried out.  The only discrepancy was
found for the very opaque core of L1495.  Luckily, a C$^{18}$O map was available for this
region (Duvert \etal 1986), which showed that the ``extra structure'' seen in the IRAS map
does indeed appear similarly in the C$^{18}$O map; the agreement is only compromised by
the presence of an embedded source which produces a spurious ``hole'' in the IRAS
map--otherwise the agreement is surprisingly good.  We conclude that the $^{13}$CO and
A$_{\rm V}$ maps are saturated, and that the IRAS 100 $\mu$m column density map probes
optical depths similar to that probed by C$^{18}$O, at least in this case.

Another indication that the IRAS column density maps presented here faithfully represent
the structure is the excellent agreement between our images and the adaptive grid/wavelet
transform optical extinction maps presented by Cambresy (1999) for Taurus, Sco-Oph, R Cr
A, and Cham-Musca.  The techniques used by Cambresy allow extinction to probe much larger
column densities, and more accurately, than previous extinction studies.\\

\begin{center}
~
\psfig{figure=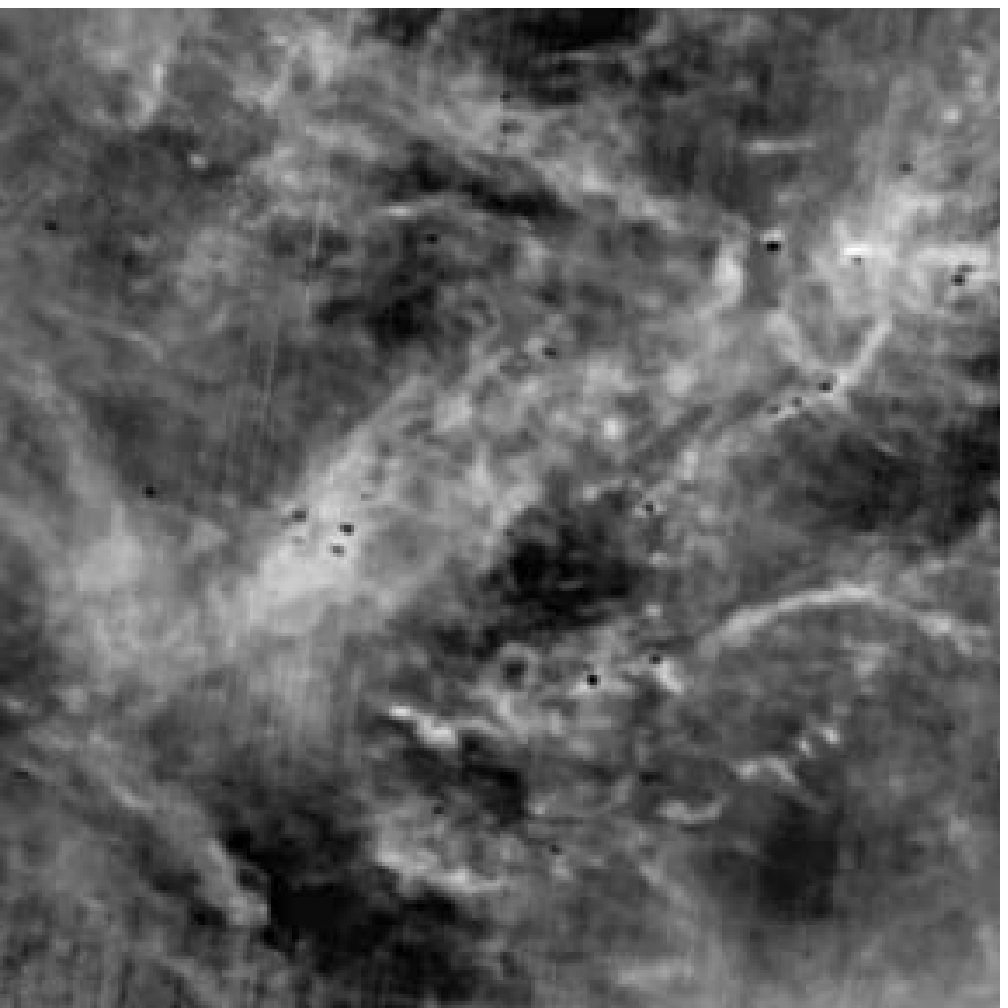}
~
\end{center}
\figcaption[fig9.smaller.eps]{100 $\mu$m column density map of Taurus, with unsharp
masking applied to enhance small scale structure. \label{fig9}}

\begin{center}
~
\psfig{figure=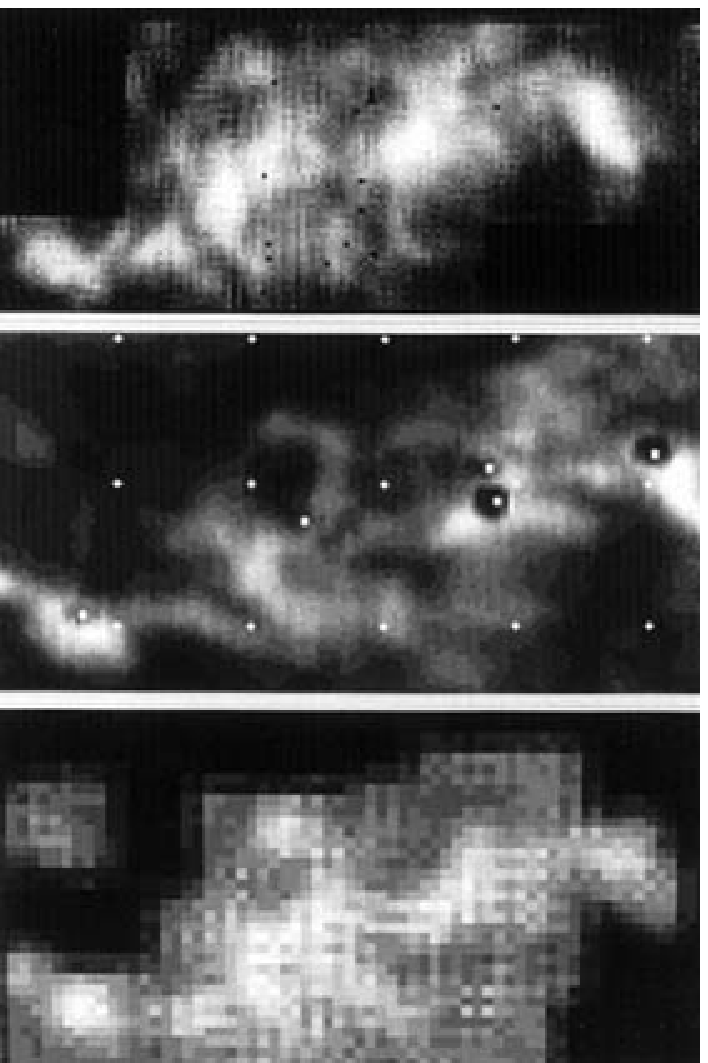}
~
\end{center}
\figcaption[fig10.smaller.eps]{Comparison of column density images for L1529 in Taurus
based on $^{13}$CO (top), N$_{100}$ (middle), and A$_{\rm V}$ (bottom).
\label{fig10}}

\section{Multiplicative Toy Models}

The \fa spectrum is ideally suited to characterizing structure which
can be created by a multiplicative process.  We present one such
process, called the p-model (see Sreenivasan 1991) or multinomial
multiplicative process (Feder 1988), to demonstrate the sensitivity
of the \fa spectrum to both scaling and geometry in images.

\begin{center}
~
\psfig{figure=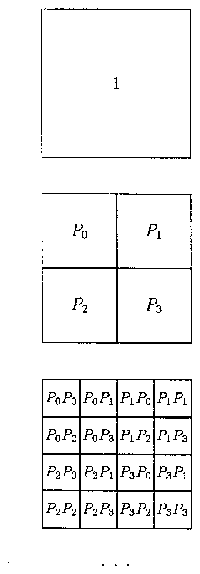}
~
\end{center}
\figcaption[fig11.smaller.eps]{The multiplicative construction process defining the toy 
models.  A square with an initially uniform measure is divided 
into four smaller squares.  In each, the original 
measure is multiplied by a constant $P_i$, where $i \epsilon \{0, 1, 2, 3\}$.
The set $\{P_i\}$ remains fixed throughout the entire construction process.
Each of the squares are then divided once again, and the construction
process is repeated a large number of times. The final
function will consist of a large number of spikes and is known to possess
multifractal scaling.  This is Sreenivisan's (1991) ``p-model.''  
The \fa curves for several multifractals generated with the 
p-model are shown in Fig. 12. \label{fig11}}

Figure 11 shows the construction process for the model.  A uniform 
square is divided into four smaller squares each receiving a 
probability $p_{\rm i}$.  Each smaller square,
in turn, is subdivided into still smaller squares, the probability in each
given by the product of $p_{\rm i}$ with the probability of the ``parent''
square.  Iterating this process produces a probability distribution
densely filled with singularities. 
Singularities with common exponents, $\alpha$, are distributed 
as homogeneous fractals; however the fractal dimension varies as a 
function of the exponent (Feder 1988).  The \fa spectrum 
provides a convenient method of characterizing both the scaling exponents
of the singularities and the fractal dimension of the ``iso-$\alpha$'' sets.

\begin{center}
~
\psfig{figure=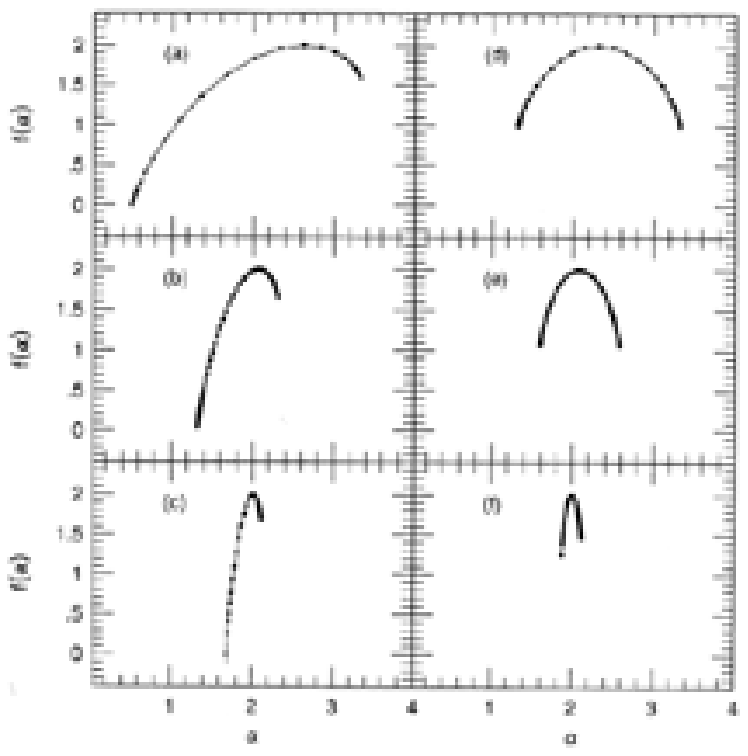}
~
\end{center}
\figcaption[fig12.smaller.eps]{The \fa singularity spectrum for the toy multiplicative models
defined in Fig. 11.  In the right-hand panels the diagonal probabilities
are set equal $P_1 = P_2$ and $P_0 = P_3 \equiv 0.5 - P_1$. 
In the left-hand panels three probabilities
are set equal $P_1 = P_2 = P_3$ with $P_0 \equiv 1 - 3 P_1$.  
In each case, the difference $|P_1 - P_0|$ increases up the page producing
stronger ``singularities.''   \label{fig12}}

Figure 12 shows the \fa spectrum for several p-models.
In each case the pattern of probabilities was kept constant
for all stages of the construction.  In the left hand column,
the largest probability is given by $p_1$ and the remaining three 
probabilities are
set equal $p_2 = p_3 = p_4 \equiv (1 - p_1)/3$.  The ratio $p_1 / p_2$ is 
decreased down the column.  All the curves peak at $f(q=0) = 2$ corresponding 
to the two-dimensionality of the images.  The asymmetry of the \fa curves 
reflects the asymmetry in the spatial distributions of the strong and
weak singularities;  the strongest singularities
are distributed as isolated points ($f(q=\infty) = 0$), whereas the weakest 
singularities are more densely distributed and
can be shown to lie on a ``Serpinski gasket'' with a fractal dimension 
$f(q = -\infty) \approx 1.58$.

The value of \al at the
$q=\infty$ point represents the strength of the strongest singularities and
decreases as the ratio $p_1/p_2$ increases.
The width of the \fa curve measures the range of the singularity strengths
present in the image.  

The curves in the right-hand column are based on p-models in which 
$p_1 = p_3$ and $p_2 = p_4 \equiv (1 - p_1)/2$.  Again, the ratio 
$p_1/p_2$ is decreased down the column.  
The strongest singularities occur where the probability distribution is 
largest, i.e. where the probabilities $p_1$ or $p_3$ have been used in each 
stage of the construction process.  Since the number of such squares at the 
$n^{th}$ construction stage scales as $n \propto L^{-1}$, where $L$ is the 
box size, the strongest singularities have a fractal dimension $D = 1$.
Similarly, the weakest singularities also have $D = 1$, producing a
symmetrical \fa curve which remains above $f = 1$.

We conclude with a cautionary remark. Figure 13 shows three 
point sets which evoke different visual impressions yet which
possess the {\em same} fractal dimension of $D = 1.0$.  
It is obvious that the fractal dimension (and therefore the \fa spectrum) 
is a severely incomplete and degenerate description of the underlying structure
and care should be taken when making inferences based on this (or any other)
structural descriptor.  A well-known similar example is the power spectrum
(and hence correlation function), which is highly degenerate (very dissimilar
structures can produce nearly-identical power spectra) because
phase information is lost in its construction.  More detailed discussion of 
the ambiguity involved in  interpreting the multifractal spectrum can be found 
in Chhabra, Jensen, and Sreenivasan (1989) and Bulajich and Perez-Pascual (1991).

\begin{center}
~
\psfig{figure=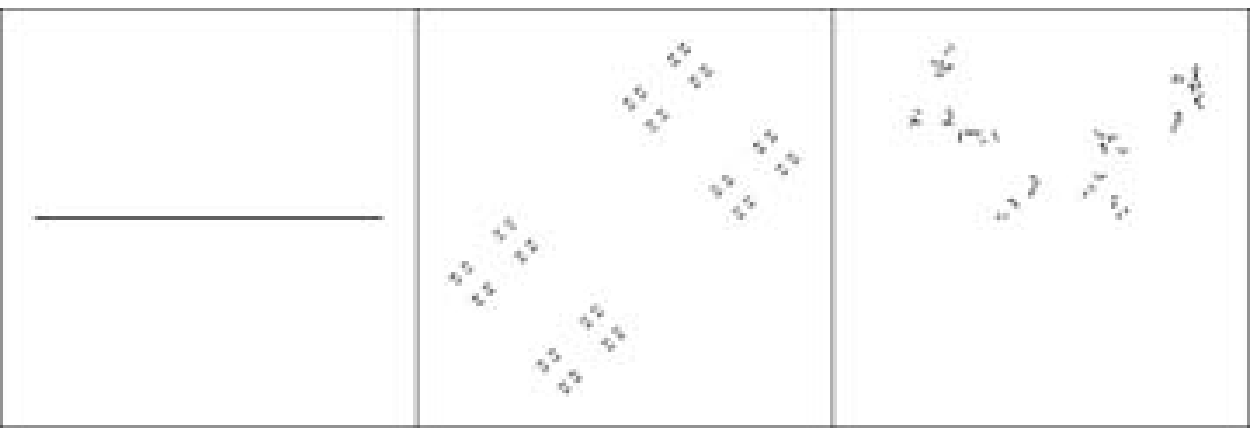}
~
\end{center}
\figcaption[fig13.smaller.eps]{Examples showing the degeneracy of the fractal dimension as
a structure descriptor.  Each point set has a fractal dimension $D = 1$.
The fractal in the center panel was created with the fixed p-model with 
$P_1 = P_2 = 0.5$ and $P_0 = P_3 = 0$ (see Fig. 11).  The ordering of the
probabilities was reversed on alternate stages of the construction process.
The fractal in the right-hand panel
was constructed using the same set of probabilities, but with random ordering
at each stage of the construction process. \label{fig13}}

\pagebreak
\begin{centering}
{\large \bf References}
\end{centering}

\begin{list}{}{}

\item \hspace{-\leftmargin} Abergel, A., Boulanger, F., Fukuii, Y., and Mizuno, A. 1995,
A\&A Supp., 111, 483

\item \hspace{-\leftmargin} Abergel, A., Boulanger, F., Mizuno, A., and Fukui, Y. 1994, ApJL,
423, L59.

\item \hspace{-\leftmargin} Adams, F. C. 1992, ApJ, 387, 572

\item \hspace{-\leftmargin} Adams, F. C. and Wiseman, J. J. 1994, 
	ApJ, 435, 693

\item \hspace{-\leftmargin} Arce, H. G. and Goodman, A. A. 1999, ApJ, 517, 264.

\item \hspace{-\leftmargin} Balafas, J. S. and Dewey, T. G. 1995, 
	Phys. Rev. E, 52, 880

\item \hspace{-\leftmargin} Bally, J. 1989, in Structure and Dynamics of the Interstellar Medium, IAU
Colloquium 120, ed. G. Tenorio-Tagle, M. Moles, and J. Melnick (Springer:
Berlin), p. 309

\item \hspace{-\leftmargin} Bally, J., Langer, W. D., Stark, A. A., and Wilson, R. W. 1987, ApJL, 312, L45

\item \hspace{-\leftmargin} Baud, B., Young, E., Beichman, C. A., \etal\ 1984, ApJ, 278, L53

\item \hspace{-\leftmargin} Bazell, D. and Desert, F. X. 1988, ApJ, 333, 353

\item \hspace{-\leftmargin} Beckwith, S.V.W., Sargent, A. I., Chini, R. S., and Gusten, R. 1990, AJ, 99, 924

\item \hspace{-\leftmargin} Beech, M. 1987, Ap. Sp. Sci., 133, 193

\item \hspace{-\leftmargin} Bhavsar, S.P. and Ling, E.N. 1988, 
	ApJL, 331, L63

\item \hspace{-\leftmargin} Bulajich, R. and P\'erez-Pascual, R. 1991, in Nonlinear Phenomena in
Fluids, Solids and Other Complex Systems, ed. P. Cordero and B. Nachtergaele
(Elsevier Sci. Publ.), p. 435

\item \hspace{-\leftmargin} Cambresy, L. 1999, A\&A, 345, 965

\item \hspace{-\leftmargin} Cernicharo, J. 1991, in The Physics of Star Formation and Early Stellar
Evolution, ed C. J. Lada and N. D. Kylafis (Dordrecht: Kluwer), p. 287

\item \hspace{-\leftmargin} Cernicharo, J. and Bachillar, R. 1984, A\&A Suppl,
58, 327

\item \hspace{-\leftmargin} Chhabra, A. and Jensen, R. V. 1989, Phys. Rev. Lett., 62, 1327

\item \hspace{-\leftmargin} Chhabra, A.B., Jensen, R. V., and Sreenivasan, K. R. 1989, 
Phys. Rev. A, 40, 4593

\item \hspace{-\leftmargin} Chhabra, A.B. and Sreenivasan, K.R. 1992, 
	Phys. Rev. Lett., 68, 2762

\item \hspace{-\leftmargin} Chen, H., Grenfell, T. G., Myers, P. C., and
	Hughes, J. D. 1997, ApJ, 478, 295

\item \hspace{-\leftmargin} Colombi, S., Bouchet, F. R., Schaeffer, R. 1992,
	A\&A, 263, 1

\item \hspace{-\leftmargin} DeGeus, E. J., de Zeew, P. T., Lub, J. 1989, A\&A, 216, 44

\item \hspace{-\leftmargin} Dickman, R. L., Horvath, M. A., and Margulis, M. 1990, ApJ, 365,
586

\item \hspace{-\leftmargin} Duvert, G., Cernicharo, J., and Baudry, A. 1986, A\&A,
164, 349

\item \hspace{-\leftmargin} Elias, J. H. 1978, ApJ, 224, 857

\item \hspace{-\leftmargin} Elmegreen, B. G. 1992 in The Galactic Interstellar Medium, SAAS FEE
Advanced Course 21, ed. D. Pfenniger and P. Bartholdi, pp. 157-274

\item \hspace{-\leftmargin} Elmegreen, B. G. 1997, ApJ, 477, 196

\item \hspace{-\leftmargin} Elmegreen, B. G. and Falgarone, E. 1996,
	ApJ, 471,816

\item \hspace{-\leftmargin} Falgarone, E., Puget, J.-L., and Perault, M. 1992, 
 	A\&A, 257, 715

\item \hspace{-\leftmargin} Falgarone, E., and Phillips, T. G. 1991, in Fragmentation of Molecular
Clouds and Star Formation, ed. E. Falgarone, F. Boulanger, and G. Duvert
(Dordrecht: Reidel), p. 119

\item \hspace{-\leftmargin} Falgarone, E., Phillips, T. G., and Walker, C. K. 1991, ApJ, 378, 186

\item \hspace{-\leftmargin} Feder, J. 1988, Fractals (New York: Plenum Press)

\item \hspace{-\leftmargin} Franco, G.A.P. 1991, A\&A, 251, 581

\item \hspace{-\leftmargin} Genzel, R. 1991, in Molecular Clouds, ed R. A. James and T. J. Millar
(Cambridge:  Cambridge Univ. Press), p. 75

\item \hspace{-\leftmargin} Graham, J. A. 1990, in Low Mass Star Formation in Southern Molecular
Clouds, ed. B. Reipurth (ESO)

\item \hspace{-\leftmargin} Gregorio-Hetem, J., Sanzovo, G. C., and Lepine, J.R.D. 1988, A\&A,
76, 347

\item \hspace{-\leftmargin} Henriksen, R. N. 1991, ApJ, 377, 500

\item \hspace{-\leftmargin} Hetem, A. and L\'epine, J.R.D. 1993, A\&A, 270, 451

\item \hspace{-\leftmargin} Heyer, M. H., Vrba, F. J., Snell, R. L., Schloerb, F. P.,
Strom, S. E., Goldsmith, P. F., and Strom, K. M. 1987, ApJ, 312, 855

\item \hspace{-\leftmargin} Hosokawa, I. and Yamamoto, K. 1990, Phys. Fluids A, 2, 889

\item \hspace{-\leftmargin} Houlahan, P. and Scalo, J. M. 1990, 
	ApJ Suppl., 72, 133

\item \hspace{-\leftmargin} Houlahan, P. and Scalo, J.M. 1992, 
	ApJ, 393, 172.

\item \hspace{-\leftmargin} Jarrett, T. H., Dickman, R. L., and Herbst, W. 1989, ApJ, 345, 881

\item \hspace{-\leftmargin} Kawamura, A., Onishi, T., Yonekura, Y., Dobashi, K., Mizumo, A.,
 Ogawa, H., \& Fukui, Y. 1999, in Star Formation 1999, ed. T. Nakamoto (Nobeyama Radio
	Observatory), p. 76

\item \hspace{-\leftmargin} Kenyon, S. J., Hartmann, L. W., Strom, K. M., and Strom, S. E. 1990,
 AJ, 99, 869

\item \hspace{-\leftmargin} Lada, C. J., Alves, J., and lada, E. A. 1999, ApJ, 512, 250

\item \hspace{-\leftmargin} Langer, W. D., Wilson, R. W., Goldsmith, P. F., 
and Beichman, L. A. 1989, ApJ, 337, 739

\item \hspace{-\leftmargin} Larson, R. 1992, MNRAS, 256, 641

\item \hspace{-\leftmargin} Lawrence, J. K., Ruzmaikin, A. A., and
	Cadavid, A. C. 1993, ApJ, 417, 805

\item \hspace{-\leftmargin} Lovejoy, S. 1982, Science, 216, 185

\item \hspace{-\leftmargin} Luo, S., Vishniac, E.T., and Martel, H. 1996, 
	ApJ, 468, 62

\item \hspace{-\leftmargin} Marraco, H. G. and Rydgren, A. E. 1981, ApJ, 86,
62

\item \hspace{-\leftmargin} Martinez, V. J., Jones, B.J.T., Dominguez-Tenreiro, R., and van de Weygaert, R.
1990, ApJ, 357, 50

\item \hspace{-\leftmargin} McCauley, J. L. 1990, 
	Phys. Reports, 189, 225

\item \hspace{-\leftmargin} Menci, N., Colafrancesco, S., and Biferale, L. 
	1993, J. de Physique I, 3, 1105

\item \hspace{-\leftmargin} Meneveau, C. and Chhabra, A. B. 1990, 
	Physica A, 164, 564

\item \hspace{-\leftmargin} Meneveau, C. and Sreenivasan, K. R. 1989, 
	Phys. Rev. Lett., 59, 1424

\item \hspace{-\leftmargin} Mott, F., Andre, P., and Neri, R. 1998, A\&A, 336, 150

\item \hspace{-\leftmargin} Muller, J. Rambak, J. P., Clausen, O. P. F., 
	and Hovig, T. 1991, in Complexity, Chaos, and Biological
	Evolution, eds. E. Mosekilde and L. Mosekilde (NY: Plenem Press), 
	p. 377

\item \hspace{-\leftmargin} Muzy, J. F., Bacry, E., and Arneodo, A. 1991, 
	Phys. Rev. Lett., 67, 3515

\item \hspace{-\leftmargin} Muzzio, F. J., Meneveaus, C., Swanson, P. D., and
	Ottino, J. M. 1992, Phys. Fluids. A, 7, 1439

\item \hspace{-\leftmargin} Nagatani, T. 1992,
	J. Phys. A, 25, L955

\item \hspace{-\leftmargin} O'Brian and Chhabra, A. B. 1992, preprint.

\item \hspace{-\leftmargin} Rossano, G. S. 1978a, AJ, 83, 234

\item \hspace{-\leftmargin} ---------\thinspace1978b, AJ, 83, 241

\item \hspace{-\leftmargin} Sargent, A. I., van Duinen, R. J., Nordh, H. L., Fridlund, G.V.M., Aalders,
J.W.G., and Beintema, D. 1983, AJ, 88, 88

\item \hspace{-\leftmargin} Scalo, J. M. 1985, in
        Protostars and Planets II, eds. D. C. Black and M. S. Mathews
        (Tucson: Univ. of Arizona Press), p. 201

\item \hspace{-\leftmargin} Scalo, J. M. 1990, in Physical Processes in Fragmentation and Star
Formation, ed. R. Capuzzo-Dolcetta \etal\ (Dordrecht: Reidel), p. 151

\item \hspace{-\leftmargin} Schwartz, R. D. 1989, in Low Mass Star Formation in Southern Molecular
Clouds, ed. B. Reipurth (ESO)

\item \hspace{-\leftmargin} Snell, R. L. 1981, ApJ, 45, 121

\item \hspace{-\leftmargin} Snell, R. L., Schoerb, F. P., Heyer, M. H. 1989, 
	ApJ, 337, 739

\item \hspace{-\leftmargin} Sreenivasan, K. R. 1991, Ann. Rev. Fluid Mech., 23,
539

\item \hspace{-\leftmargin} Sreenivasan, K. R. 1996, 
	in Research Trends in Fluid Dynamics, eds. J. L. Lumley,
	A. Acrivos, L. G. Leal, and S. Leibovich (Woodbury, N.Y.: AIP Press),
	p. 263

\item \hspace{-\leftmargin} Stutzki, J., Genzel, R., Graf, U., Harris, A., Sternberg, A., and Gusten, R.
1991, in Fragmentation of Molecular
Clouds and Star Formation, ed. E. Falgarone, F. Boulanger, and G. Duvert
(Dordrecht: Reidel), p. 235

\item \hspace{-\leftmargin} Sylos Labini, F. and Pietronero, L. 1996,
	ApJ, 469, 26

\item \hspace{-\leftmargin} Tachihara, K., Mizuno, A., \& Fukui, Y. 1999, preprint

\item \hspace{-\leftmargin} Takayasu, H. and Suzuki, J. 1991, J. Phys. A., 24,
L1309

\item \hspace{-\leftmargin} Veneziano, D., Moglen, G. E., and Bras, R. L. 
	1995, Phys. Rev. E, 52, 1387

\item \hspace{-\leftmargin} Verter, F. and Rickard, L. J. 1998, AJ, 115, 745

\item \hspace{-\leftmargin} Vogelaar, M. G. R. and Wakker, B. P. 1994, 
	A\&A, 291, 557

\item \hspace{-\leftmargin} Vogelaar, M. G. R., Wakker, B. P., and Schwarz, U. J. 1991, in Fragmentation of Molecular
 Clouds and Star Formation, ed. E. Falgarone, F. Boulanger, and G. Duvert
 (Dordrecht: Reidel), p. 508

\item \hspace{-\leftmargin} Wakker, B. P. 1990, Ph.D. dissertation, Univ. of Groningen

\item \hspace{-\leftmargin} Walter, F. M. and Boyd, W. 1991, ApJ, 370, 318

\item \hspace{-\leftmargin} Westpfahl, D. J., Coleman, P. H., Alexander, J., \& Tongue, T.
 1999, ApJ, 117, 868.

\item \hspace{-\leftmargin} Whittet, D.C.B., Kirrane, T. M., Kilkenny, D., Oates, A. P.,
Watson, F. G., and King, D. J. 19487, MNRAS, 224, 497

\item \hspace{-\leftmargin} Wilking, B. A., Schwartz, R. D., and Blackwell, J. H. 1987,
AJ, 94, 106

\item \hspace{-\leftmargin} Wilson, T. L. and Walmsley, C. M. 1989, 
	A\&A Rev, 1, 141

\item \hspace{-\leftmargin} Xie, T., Goldsmith, P. F., and Zhou, W. 1991, ApJ, 
371, L81

\item \hspace{-\leftmargin} Yamaguti, M. and Prado, C. P. C. 1995, 
	Phys. Lett. A, 206, 318

\item \hspace{-\leftmargin} Yonekura, Y., Mizuno, N., Saito, H., Mizuno, A., Ogawa, H.,
 \& Fukui, Y. 2000, PASJapan, in press

\item \hspace{-\leftmargin} Zinnecker, H., McCaughrean, M. J., and Wilking, 
	B. A. 1993, in Protostars and Planets III, 
	eds. E. H. Levy, J. I. Lunine (U. of Arizona Press: Tucson), p. 429
\end{list}

\clearpage

\begin{deluxetable}{lrrrrr}
\tablenum{1}
\tablewidth{0pt}
\tablecaption{Properties of the Regions Studied}
\tablehead{
	\colhead{Subregion}           	& \colhead{R. A.}      &
	\colhead{Dec.}          	& \colhead{\# Res. Elem.}  &
	\colhead{Dist. (pc)}          	& \colhead{\# YSO/pc$^{2}$}
}
\startdata
Tau       & $4^{h} 30'$   & $+26^{\circ}.0$  &  $3.2\times10^{4}$ & $140^{a}$ & $0.3^{b}$  \\ 
R CrA W   & $19^{h} 04'$  & $-37^{\circ}.0$  &  $1.0\times10^{4}$ & $130^{c}$ & $3.0^{d}$  \\
R CrA E   & $19^{h} 32'$  & $-37^{\circ}.0$  &  $2.6\times10^{4}$ & $130^{c}$ & $0.0^{d}$  \\ 
Cham 1    & $11^{h} 00'$  & $-77^{\circ}.0$  &  $1.7\times10^{4}$ & $140^{e}$ & $3.5^{f}$  \\
Cham 2    & $12^{h} 40'$  & $-78^{\circ}.0$  &  $1.1\times10^{4}$ & $140^{e}$ & $2.0^{f}$  \\ 
Cham 3    & $12^{h} 20'$  & $-80^{\circ}.0$  &  $1.4\times10^{4}$ & $140^{e}$ & $0.0^{f}$  \\
Musca     & $12^{h} 20'$  & $-71^{\circ}.0$  &  $1.5\times10^{4}$ & $140^{e}$ & $0.1^{f}$  \\
$\rho$ Oph & $16^{h} 26'$ & $-26^{\circ}.0$  &  $5.6\times10^{4}$ & $125^{g}$ & $23.0^{h}$ \\
Oph E     & $16^{h} 50'$  & $-24^{\circ}.0$  &  $2.0\times10^{4}$ & $125^{g}$ & \nodata \\
Oph W     & $15^{h} 40'$  & $-27^{\circ}.0$  &  $3.6\times10^{4}$ & $125^{g}$ & \nodata \\
Oph U     & $16^{h} 26'$  & $-19^{\circ}.0$  &  $2.2\times10^{4}$ & $125^{g}$ & \nodata \\
Oph N     & $16^{h} 40'$  & $-10^{\circ}.0$  &  $1.0\times10^{5}$ & $125^{g}$ & \nodata \\
L 134     & $15^{h} 56'$  & $-5^{\circ}.0$   &  $2.4\times10^{4}$ & $160^{i}$ & $0^{j}$ 
\tablerefs{
(a) Elias 1978; (b) Beckwith {\em et al.} 1990, Kenyon {\em et al.} 1990; 
(c) Marraco and Rydgren 1981; (d) Graham 1990; 
(e) Whittet {\em et al.} 1987, Franco 1991;  (f) Schwartz 1989, Baud 1984;
(g) deGeus {\em et al.} 1989; (h) Wilking {\em et al.} 1989; (i) Snell 1981;
(j) Sargent {\em et al.} 1983.}
\enddata
\end{deluxetable}

\end{document}